\documentclass[fleqn,12pt,twoside]{article}
\usepackage{espcrc1}

\usepackage{graphicx}
\usepackage[figuresright]{rotating}

\newcommand{\AmS}{{\protect\the\textfont2
A\kern-.1667em\lower.5ex\hbox{M}\kern-.125emS}}

\title{ 
Coulomb three-body systems and charge transfer collisions in a
configuration-space
approach}

\author{Renat A. Sultanov\thanks{Supported by
NSF  Grant  ATM-0205199}\address[MCSD]{Department of Chemistry, 
University of Nevada,
Las Vegas,
Nevada 89154, USA}
and Sadhan K. Adhikari\thanks{Supported in part by the
CNPq of Brazil}\address{Instituto de F\'isica
Te\'orica, Universidade Estadual Paulista, 01405 S\~ao Paulo, SP, Brazil}}
       
\begin{document}

\maketitle

\begin{abstract}Some
low-energy three-body muon- and electron-transfer  processes are
considered within the Faddeev-Hahn formulation using two-, six-, and
ten-state close-coupling approximation. We test our approach in 
bound-state problems  for systems  H$_2^+$ and ($\mu^-$dd)$^+$ within six-
and ten-state schemes, where $d$ is a deuteron and $\mu$ a muon. We
present results in the
six-state model for
muonium formation from hydrogen. We also present results for muon-transfer
rates from  muonic
hydrogen isotopes to bare nuclei S$^{16}$ and Ar$^{18}$ in reasonable
agreement with experiment. For these heavier targets a polarization
potential is included and  Coulomb potentials are treated exactly without
approximation or cut off.
\end{abstract}

\vspace{6mm} 

A theoretical description of low-energy three-body rearrangement reaction 
with the exchange of a heavy particle invalidates the simplifying
Born-Oppenheimer approximation and requires a full quantum
treatment. Here we study several three-body charge-transfer reactions and
bound-state
problems 
using the
Faddeev-Hahn formulation \cite{2} where the full wave function is broken
into two
components. In the case of three charged particles composed of positive
and negative charges there are at most two asymptotic two-cluster bound
configurations. The two components of the wave function carry
the asymptotic boundary condition in these two channels. For numerical
treatment we employ two-, six-, and ten-state close-coupling schemes. 

After testing the approach to the bound states H$_2^+$ and
($\mu^-$dd)$^+$   we consider
charge-transfer reactions.
Hydrogen has three isotopes: protium H, deuterium D, and tritium T. It may
be an important test for any formulation  to predict reaction rates as a
function of 
isotopes.  Muonium (Mu) is composed
of a positive muon $\mu^+$ and an electron and can be treated as an
ultra-light hydrogenic isotope compared to protium, deuterium and tritium,
because the proton is about nine times heavier than the muon.
Because of
differences in mass,  muonium formation  can demonstrate rather different
dynamical
properties than the reactions of other isotopes of hydrogen. 
For this reason   we consider the reaction for muonium production in
muon-hydrogen collision: $\mu^{+} + \mbox{H}_{1s} \rightarrow
\mbox{Mu}_{1s} +$
H$^+$.

Due to interest in the possibility of muon-catalyzed fusion of the
hydrogen isotopes, a study of muon transfer from hydrogen isotopes to
nuclei X$^{Z+}$ with charge $Z$ is of great interest. There have been
several experimental measurements of muon transfer from muonic hydrogen
isotopes $^1$H$_\mu$ and  $^2$H$_\mu$ to bare nuclei H, He, C, O, Ne, S,
Ar,
Kr, Xe etc. denoted by  (H$_\mu$)$_{1s}$  + X$^{Z+}$  $\to$
(X$_\mu$)$^{(Z-1)+}$ + H$^+$. The rates increase by few orders of
magnitude as one moves from the light nuclei H, He to the heavy
ones. These
reactions are  very relevant as a small contamination of these heavier
elements will significantly influence the process of muon-catalyzed 
fusion of hydrogen isotopes. Muon-transfer reaction with these nuclei has very strong
final-state Coulomb  interaction    and is  a real
challenge to any theoretical formulation.
 Here we report results for muon transfer rates for
S$^{16+}$
and A$^{18+}$. 

The well-known Faddeev equations in momentum space \cite{2} are
supposed to be the most adequate way of considering
rearrangement scattering problems at
low energies.  However, a direct
application of the integral Faddeev equations to Coulombic three-body
systems
is impossible due to long-range character of the Coulomb interaction.
The Coulomb interaction  brings in significant singularities in the
kernels of
these equations.

Recently, we formulated a few-body quantum-mechanical description of
direct charge-transfer reaction from hydrogen
to different nuclei using Faddeev-Hahn-type equations in the close-coupling 
approximation \cite{3}.
In this approach the three-body wave function is taken as
$\Psi =  \Psi_1 (\vec r_{23},\vec \rho_1) +
\Psi_2 (\vec r_{13},\vec \rho_2),
$
where $\vec r_i, i=1,2,3$ is the position vector of particle $i$, $\vec
r_{ij}= \vec r_i - \vec r_j$, and $\vec \rho_k$ is the relative position
of $k$ from the center of mass (cm) of $i$ and $j$. Here 
$ \Psi_1 (\vec r_{23}, \vec \rho_1)$ is quadratically integrable
over the variable $\vec r_{23}$, and $\Psi_2 (\vec r_{13},\vec
\rho_2)$ over the variable $\vec r_{13}$. These components satisfy the
Faddeev-Hahn equations \cite{2}
\begin{eqnarray}\label{eq:FHo}
(E-H_0-V_{23}-U_P)\Psi_1 (\vec r_{23}, \vec \rho_1)&=&
(V_{23}+V_{12}-U_C)
\Psi_2 (\vec r_{13}, \vec \rho_2),\\
(E-H_0-V_{13}-U_C)\Psi_2 (\vec r_{13}, \vec \rho_2)&=&
(V_{13} +V_{12}-U_P)
\Psi_1 (\vec r_{23}, \vec \rho_1).
\label{eq:FH}
\end{eqnarray}
Here $H_0$ is the total kinetic energy operator, $V_{ij}$
the  pair Coulomb potential,
$E$ the total cm  energy, $U_P= 9 Z^2/(4\rho^4)$ the polarization
potential (included with a short-range cut off  for only muon transfer to
Ar
and S), $U_C$  the
final-state Coulomb interaction between 2 and
bound state (13). The close-coupling equations based on these equations
are projected on the total angular momentum $J=0$ appropriate at low
energies and then solved numerically by discretization.

\vskip -0.7 cm
\begin{figure}[hb]
\centering
\includegraphics[width=0.53\linewidth]{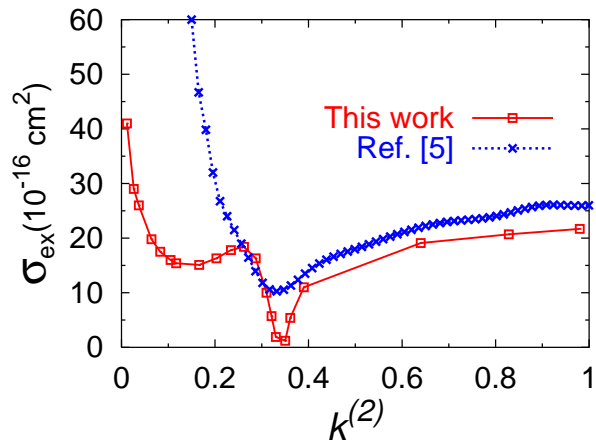}     
\vskip -0.8 cm
\caption{Cross sections $\sigma_{ex}(k^{(2)})$
of the reaction $\mu^{+} + \mbox{H}_{1s} \rightarrow \mbox{Mu}_{1s} +
$ H$^+$; (red) line with square: present results,
(blue) line with $\times$: Ref. \cite{4}}\label{fig1}
\end{figure}
\vskip -.7cm

For the binding energy  of H$_2^+$, the present 6-
[2$\times$(1s-2s-2p)] and
the 10-state
[2$\times$(1s-2s-2p-3s-3p)] approximations to the close-coupling expansion
produce 0.66 au  and 0.62 au,
respectively,
compared to the accurate
result of 0.597139 
au \cite{moss}. For the spectral levels of the muonic molecular ion $(\mu
^-dd)^+$
 the six- and ten-state models lead to $-326.4$ eV and $-325.2$ eV,
respectively,
for the ground-state energy compared to the accurate variational result of 
$-325.0735$ eV \cite{monk}. The corresponding numbers for the lowest
excited
state are
$-33.1$ eV, $-35.0$ eV, and $-35.8444$ eV \cite{monk}, respectively.

  Cross sections $\sigma_{ex}$ for muon production are presented in
Fig. 1 as function of $k^{(2)} =3.7 \times \sqrt {E(\mbox{eV})}$
together
with the theoretical results of work \cite{4}. There is substantial
difference between the two calculations at low energies.  
We found that the contribution of $p$-waves in sub-systems
is reasonable and ranges up to $20 - 40\%$.

\begin{table}
{Table 1. Partial $\lambda^{\scriptsize \mbox{tr}}_{1s\rightarrow
nl}/10^{10}$ s$^{-1}$ and total  $\lambda^{\scriptsize
\mbox{tr}}_{tot}/10^{10}$ s$^{-1}$ muon transfer rates 
reduced to liquid hydrogen density $N_0= 4.25 \times 10^{22}$ cm$^{-3}$
from muonic hydrogen to Ar$^{18+}$ (at 0.04 eV) and S$^{16+}$
(at 0.01 eV) together with some
experimental results \cite{3a}.
}
\begin{tabular}{lcccccccclccccccc}
\hline
\multicolumn{1}{l}{System}                &
\multicolumn{1}{c}{ }                     &
\multicolumn{2}{c}{Without polarization} &
\multicolumn{2}{c}{With polarization} &
\multicolumn{2}{c}{Experiment}\\
\multicolumn{1}{l}{}              &
\multicolumn{1}{l}{$(nl)$}                &
\multicolumn{1}{c}{$\lambda^{\scriptsize \mbox{tr}}_{1s\rightarrow nl}$} 
&
\multicolumn{1}{c}{$\lambda_{\scriptsize
\mbox{tot}}^{\scriptsize \mbox{tr}}$} &
\multicolumn{1}{c}{$\lambda^{\scriptsize \mbox{tr}}_{1s\rightarrow nl}$} 
&
\multicolumn{1}{c}{$\lambda_{\scriptsize
\mbox{tot}}^{\scriptsize \mbox{tr}}$} & &
\multicolumn{3}{c}{$\lambda_
{\scriptsize \mbox{tot}}^{\scriptsize \mbox{tr}}$}  &
  \\ \hline
 $^1$H$_\mu$+Ar 
& $ 10s $ & $ 5.0 \pm 0.2 $ & $ $ & $ 8.1 \pm 0.2 $
& $  $ & $ 12$ & 14.6\\
$ $
& $ 10p $ & $ 3.9 \pm 0.2 $ & $ 8.9 \pm 0.4 $ & $ 4.8 \pm 0.2 $
& $ 12.9 \pm 0.4 $ &  35 & 16.3  \\
\hline
 $^2$H$_\mu$+Ar 
& $ 10s $ & $ 1.3 \pm 0.1 $ & $ $ & $ 3.4 \pm 0.2 $
& $  $ & $ $ & $ $\\
$ $
& $ 10p $ & $ 0.9 \pm 0.1 $ & $ 2.2 \pm 0.2 $ & $ 1.9 \pm 0.2 $
& $ 5.3 \pm 0.4 $ & $ 8.6 $ & $ 9.4 $  \\
\hline
 $^1$H$_\mu$+S 
& $ 9s $ & $6.5$ & $ $ & $ 8.2 \pm 0.2 $
& $  $ & $ $ & $ $\\
$ $
& $ 9p $ & $ 3.1$ & $ 9.6 $ & $ 3.8 \pm 0.2 $
& $ 12.0 \pm 0.4 $ & $ 8.9 $ & $  $  \\
\hline
 $^2$H$_\mu$+S 
& $ 9s $ & $ 6.8 $ & $ $ & $ 7.9 \pm 0.2 $
& $  $ & $ $ & $ $\\
$ $
& $ 9p $ & $ 4.0 $ & $ 10.8 $ & $4.8 \pm 0.2 $
& $ 12.7 \pm 0.4 $ & $ 11 $ & $  $  \\
\hline
\end{tabular}
\end{table}

Finally, we present the results of muon transfer from H isotopes to Ar and
S. In this case the transfer takes place mostly to  two levels of
the muonic nuclei and we consider only these two levels in the present
calculation. The muon is captured predominantly in the 10$s$ and  $10p$
states of Ar and $9s$ and $9p$ states of S \cite{ann}. We compare the
results in Table 1 with
available experiments quoted in Ref. \cite{3a}. The correct order of
magnitude for the rates is
already produced without a polarization potential. A
previous calculation of muon-transfer
rates
to lighter
nuclei (He, C, O)
\cite{3}  is  also in agreement with experiment and a very recent
calculation \cite{DL}.
From this and previous studies \cite{3} we conclude that the present
formulation is
equally
suitable for  bound states as well as three-body charge transfer
reactions involving large final-state Coulomb interaction.

\end{document}